\documentclass[11pt]{article}
\usepackage{cospar}
\usepackage{url}

\usepackage{graphicx}
\usepackage[figuresright]{rotating}

\title{VLT observations of the solitary millisecond pulsar PSR J2124$-$3358}

\author{Roberto P. Mignani\address{European Southern Observatory, Karl-Schwarzschild-Str.\ 2, D85740, Garching, Germany} and Werner Becker\address{Max--Planck--Institut f\"ur Extraterrestrische Physik, Giessenbachstra{\ss}e, 85741 Garching, Germany}}

\begin{document}

\maketitle

\begin{abstract}
About 100 millisecond (ms) pulsars have been identified in the Galaxy,
  and only $\approx 10\%$ of  them are solitary, i.e.~without a binary
  companion.  Nothing  is known on the optical  emission properties of
  millisecond  pulsars.  Observations  of {\em  solitary}  millisecond
  pulsars are  the only way  to detect their faint  optical radiation,
  otherwise buried by  the brighter white dwarf companion.   As in the
  case  of  solitary,  non  millisecond pulsars,  an  X-ray  detection
  represents the first step for a follow-up identification campaign in
  the  optical.  Among  the  X-ray detected  millisecond pulsars,  PSR
  J2124$-$3358 stands  out as an ideal  case because it  is very close
  ($\le 270$  pc) and little absorbed.  Here, we report  on recent VLT
  observations of the PSR  J2124$-$3358 aimed at the identification of
  its optical  counterpart.  No optical  emission from the  pulsar has
  been detected down to a limiting flux of $V \sim 27.8$.
\end{abstract}

\section*{Introduction}

The about 1500 radio pulsars  known to date are interpreted as rapidly
 spinning and strongly magnetized neutron stars which are radiating at
 the expense  of their  rotational energy.  Millisecond
 pulsars  form a  separate group  among the  rotation-powered pulsars.
 They are distinguished by their  small spin periods ($\le 20$ ms) and
 their high  spin stability  ($\dot P \approx  10^{-18} -  10^{-21}$ s
 s$^{-1}$),  with  corresponding   spin-down  ages  $P/2{\dot  P}$  of
 typically $10^9-10^{10}$ years.  Only  $\approx 20\%$ of the galactic
 millisecond pulsars are solitary,  including PSR B1257+12 which is in
 a planetary system. The rest are in binaries, usually with a low-mass
 white  dwarf   companion.   Optical  observations   of  {\em  binary}
 millisecond pulsars have so far  only allowed to detect the companion
 star (see, e.g., Danziger et  al.~1995; Lundgreen et al.~1996) and to
 constrain the global  evolution of the binary systems.  In no case it
 was possible to detect the  fainter optical emission from the neutron
 star which is buried by the brighter companion.  Optical observations
 of solitary millisecond  pulsars thus are the only  chance to explore
 the optical  emission properties of  these peculiar objects.   \\ The
 optical  emission of  solitary  millisecond pulsars  can be  ascribed
 either to  magnetospheric or thermal radiation from  the neutron star
 surface.   However,  unless  some  reheating mechanism  is  at  work,
 optical thermal  radiation is  expected to be  virtually undetectable
 for old ($\approx 10^{9}$ yrs),  cooled off, neutron stars and magnetospheric
 radiation appears  the most likely  emission process. From  the model
 developed by  Pacini (1971), the magnetospheric  emission is expected
 to depend on  the pulsar's spin parameters according  to the relation
 $L_{opt}  \propto B^{4}  P^{-10}$. Although  valid for  young pulsars
 such a dependance, as well as any breakdown with the pulsar's age, is
 still  unclear for  older neutron  stars. Detecting  optical emission
 from millisecond pulsars, i.e. short period and low $\dot P$ objects,
 would  thus   be  of  paramount   importance  to  define   a  general
 framework.  \\  As  for  classical,  i.e.,  isolated  non-millisecond
 pulsars, a  detection at soft X-ray energies  represents an important
 indicator   that  the   object  is   likely  detectable   at  optical
 wavelength. Currently, soft X-ray  emission have been discovered from
 a number of galactic solitary  millisecond pulsars but only for a few
 of  them X-ray  pulses have  been detected  (see Becker  \& Tr\"umper
 1999; Becker \& Aschenbach 2002).

\section*{PSR J2124$-$3358}

One of the  most interesting objects is PSR  J2124$-$3358.  The pulsar
 was identified  in radio  during the Parkes  436 MHz  southern survey
 (Bailes et al.  1997).  Its measured $\dot P$ of $1.096 ~ 10^{-20}$ ~
 s~s$^{-1}$ gives  an age  of 7.2 billion  years, a magnetic  field of
 $2.3 ~  10^{8}$ G  and a  rotational energy loss  of $3.5  ~ 10^{33}$
 ergs~s$^{-1}$.   PSR  J2124$-$3358   was  discovered  in  soft  X-ray
 energies  during  a recent  study  of  galactic isolated  millisecond
 pulsars using  the {\sl ROSAT}  HRI (Becker \& Tr\"umper  1999).  Its
 X-ray emission  was found to  be pulsed at  the radio period,  with a
 sharp  double-peaked pulse  profile.  The  shape of  the  X-ray pulse
 profile was considered a strong evidence that the X-ray emission from
 this  millisecond   pulsar  is  dominated   by  non-thermal  emission
 (cf.~Becker  \&  Tr\"umper   1999).   Indeed,  the  X-ray  lightcurve
 exhibits  a double  peak structure  remarkably similar  to  the pulse
 profile observed in the radio domain at 436 MHz (Bailes et al.  1997;
 Becker \& Tr\"umper  1999).  Figure 1 shows a  comparison between the
 PSR J2124$-$3358  pulse profiles  as oberved in  the X-ray  and radio
 bands.  The  interpretation in  terms of magnetospheric  emission has
 been  later supported  by  ASCA observations  which  showed that  the
 pulsar spectrum in the 0.5-10 keV interval is compatible with a power
 law (Sakurai  et al. 2001). \\  PSR J2124$-$3358 is  therefore one of
 the  most promising  millisecond pulsars  for optical  studies  as it
 turns out  to be  one of the  brightest solitary  millisecond pulsars
 detected  at X-ray  energies.  Its  close  distance of  about 270  pc
 (Cordes  \&  Lazio 2002)  together  with  its  small Hydrogen  column
 absorption   of  $\sim   2-5  ~10^{20}\;\mbox{cm}^{-2}$   (Becker  \&
 Tr\"umper 1999) also make this object a natural target for subsequent
 optical observations.

 \begin{figure}            
\begin{center}
 \includegraphics[width=90mm]{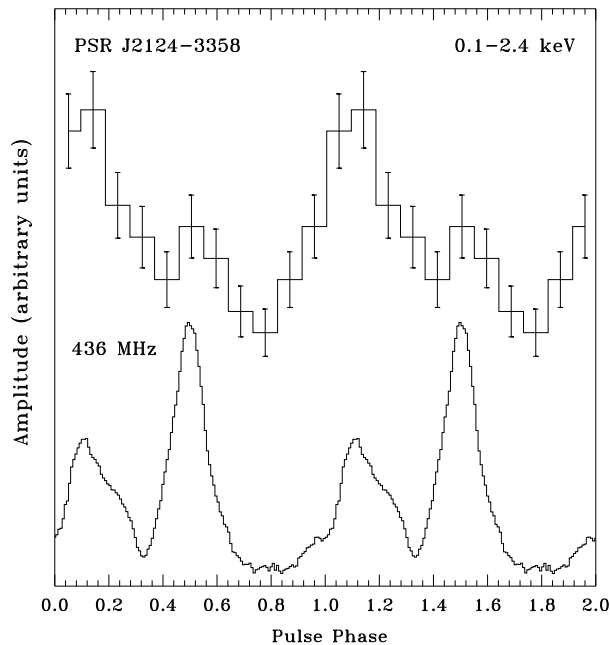}
\end{center}
\caption{Comparison between the X-ray  and radio pulse profiles of PSR
J2124$-$3358.   The  gross  similarities  in the  pulse  structure  is
obvious and indicates that the emission process which powers the X-ray
pulsations is closely related to the radio emission ones. This clearly
suggests a magnetospheric origin also for the X-ray emission.}
\end{figure}

\section*{Observations}

The  first optical  observations  of the  PSR~J2124$-$3358 field  were
 performed in July 1998, shortly  after the detection of the pulsar in
 X-rays.  Observations were  performed with the ESO NTT  using the ESO
 Multi Mode  Instrument (EMMI)  in its Blue  Arm configuration  with a
 pixel size of  0.37" (field of view $6.2' \times  6.2'$).  A total of
 six exposures (1200s each) were acquired through a $B$ filter. Single
 images were reduced using standard MIDAS recipes and combined using a
 median filter algorithm. Photometric calibration was computed through
 short exposures  of Landolt's  fields.  Unfortunately, since  all the
 exposures  of  the  PSR  J2124$-$3358  field  were  affected  by  bad
 atmospheric conditions and  by a seeing constantly above  1.5" it was
 not possible  to reach a detection  limit fainter than  $B \simeq 25$
 (Becker \&  Mignani, unpublished). \\
 The field  of PSR J2124$-$3358
 was observed  again between August  and September 2001 using  the ESO
 VLT/Antu  telescope  operated  in   Service  Mode  from  the  Paranal
 Observatory (Becker  et al. 2002).  Imaging was  performed using the
 FOcal  Reducer and  Spectrograph  1 (FORS1)  camera  operated in  its
 standard  resolution  mode,  with  a   pixel  size  of  0.2''  and  a
 corresponding field of view of $6.8' \times 6.8'$.  Images were taken
 through  the  $U$  ($\lambda=3660$\AA,  $\Delta\lambda=360$\AA),  $B$
 ($\lambda=4290$\AA,       $\Delta\lambda=880$\AA)       and       $V$
 ($\lambda=5540$\AA, $\Delta\lambda=1110.5$\AA) passbands to allow for
 a significant  spectral coverage.  For  each filter, sequences  of 12
 repeated exposures were obtained for a total integration time of 9360
 s in  $U$ and 6000  s in  $B$ and $V$.   The detailed summary  of the
 observations is listed in Table  1. To allow for relative adjustments
 in the flux  calibration, one exposure of each  sequence was obtained
 under photometric  conditions.  The pulsar field  was always observed
 close to the minimum airmass  with average values of 1.10 ($U$), 1.09
 ($B$)  and 1.03 ($V$).   Seeing conditions  across all  the exposures
 varied  between $\sim 0.6"$  and $\sim  1.0"$. \\  
Standard reduction
 steps were  applied through the FORS1 image  reduction pipeline.  For
 each night, master  bias and sky flats were used for debiassing
 and flatfielding.  Flux calibration was computed using the extinction
 and  color-corrected photometric  zero-points routinely  computed for
 each night.  For each  passband, sequences of repeated exposures were
 finally  co-added by  applying a  median filter  algorithm  to reject
 cosmic ray hits.

\begin{table}[h]
\begin{center}
\begin{tabular}{lccccc} \hline \hline
Date & Filter &   No.\ of exp. & Exposure(s) & seeing (") & airmass \\ \hline
2001 August 12   & $U$ & 3 & 780 & 1.0" & 1.33 \\ \hline
2001 August 13   & $U$ & 9 & 780 & 0.8" & 1.026\\
                 & $B$ & 1 & 500 & 0.8" & 1.034\\  
		 & $V$ & 1 & 500 & 0.8" & 1.028\\ \hline
2001 Sep 14      & $B$ & 6 & 500 & 0.58" & 1.134 \\
		 & $V$ & 6 & 500 & 0.63" & 1.046 \\\hline
2001 Sep 17      & $B$ & 5 & 500 & 0.52" & 1.054\\ \hline
2001 Sep 18      & $V$ & 5 & 500 & 0.96" & 1.017\\ \hline \hline
\end{tabular}
\end{center}
\caption{Summary   of the  optical observations  of   the field of PSR
J2124$-$3358  obtained with the FORS1 instrument  at VLT/Antu. All the
observations were taken  with the same  instrument configuration.  For
each observation, the columns give the observing date, the filter, the
number  of exposures, the integration  time per  exposure, the average
seeing conditions,   and  the  average  airmass  during  the  exposure
sequence.  } 
\end{table}

\section*{Results}

Since the first hint to claim the optical identification of the pulsar
comes from  a positional coincidence with a  potential counterpart, an
accurate astrometry of the field is a key issue.  In order to register
the  very  accurate  (down  to few  milliarcseconds)  pulsar  position
(Gaensler,  Jones and  Stappers 2002)  on  the FORS1  images, we  have
recomputed the FORS1 astrometry using  as a reference the positions of
stars  selected  from the  Guide  Star  Catalogue  II, which  have  an
intrinsic  absolute  astrometric  accuracy  of  $\approx  0.35''$  per
coordinate.  After identifying a  number of well suited GSC-II objects
in   the  co-added   $V$-band  image,   the   pixel-to-sky  coordinate
transformation    has     been    computed    using     the    program
ASTROM\footnote{http://star-www.rl.ac.uk/Software/software.htm}.    The
final precision  on the astrometric fit  was $\sim 0.09''$  in both RA
and Dec.  To this value we then  added an error of $0.17''$ due to the
propagation   of  the   intrinsic  absolute   errors  on   the  GSC-II
coordinates. 

\begin{figure}          
\includegraphics[width=85mm]{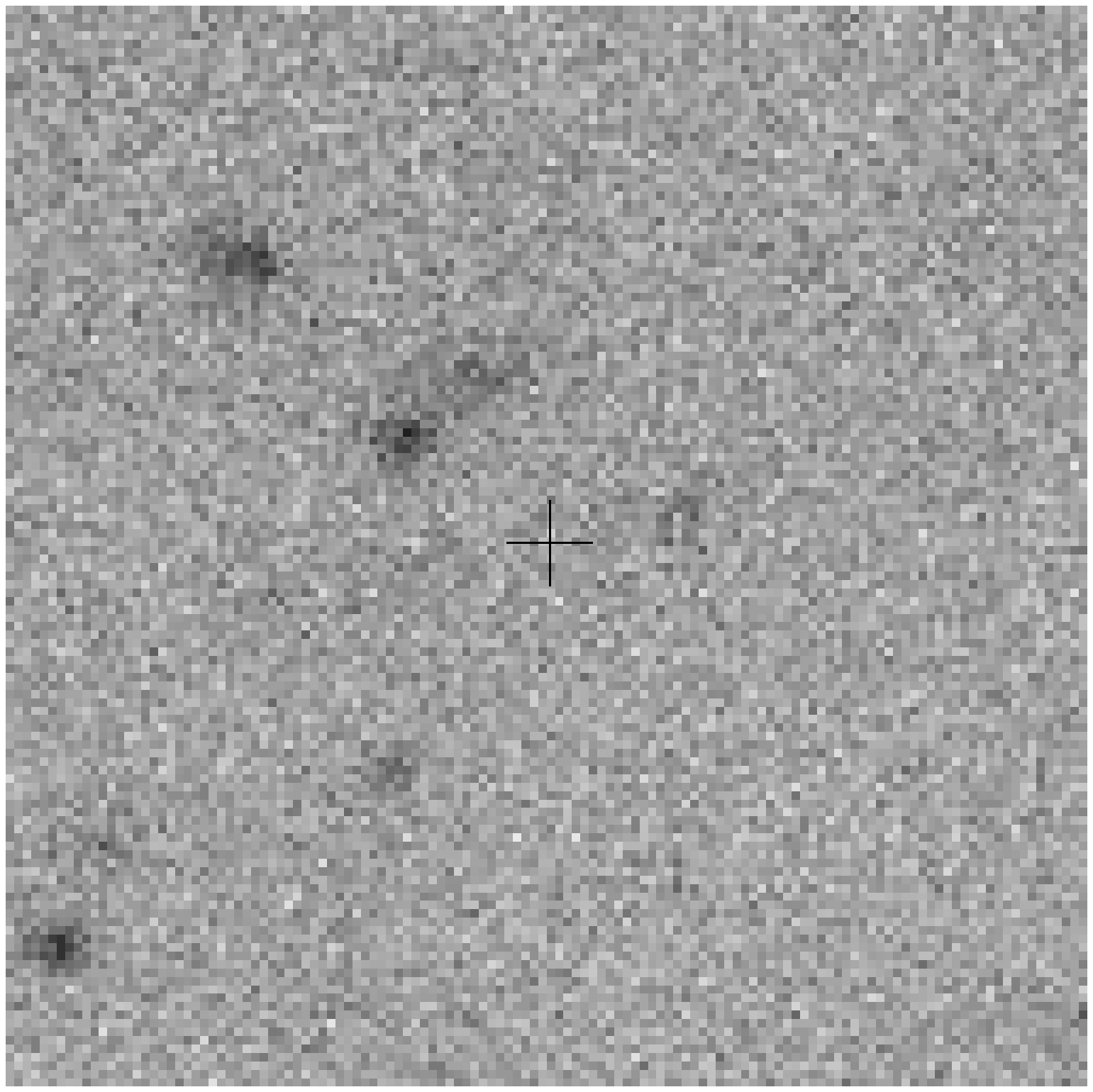}
\includegraphics[width=85mm]{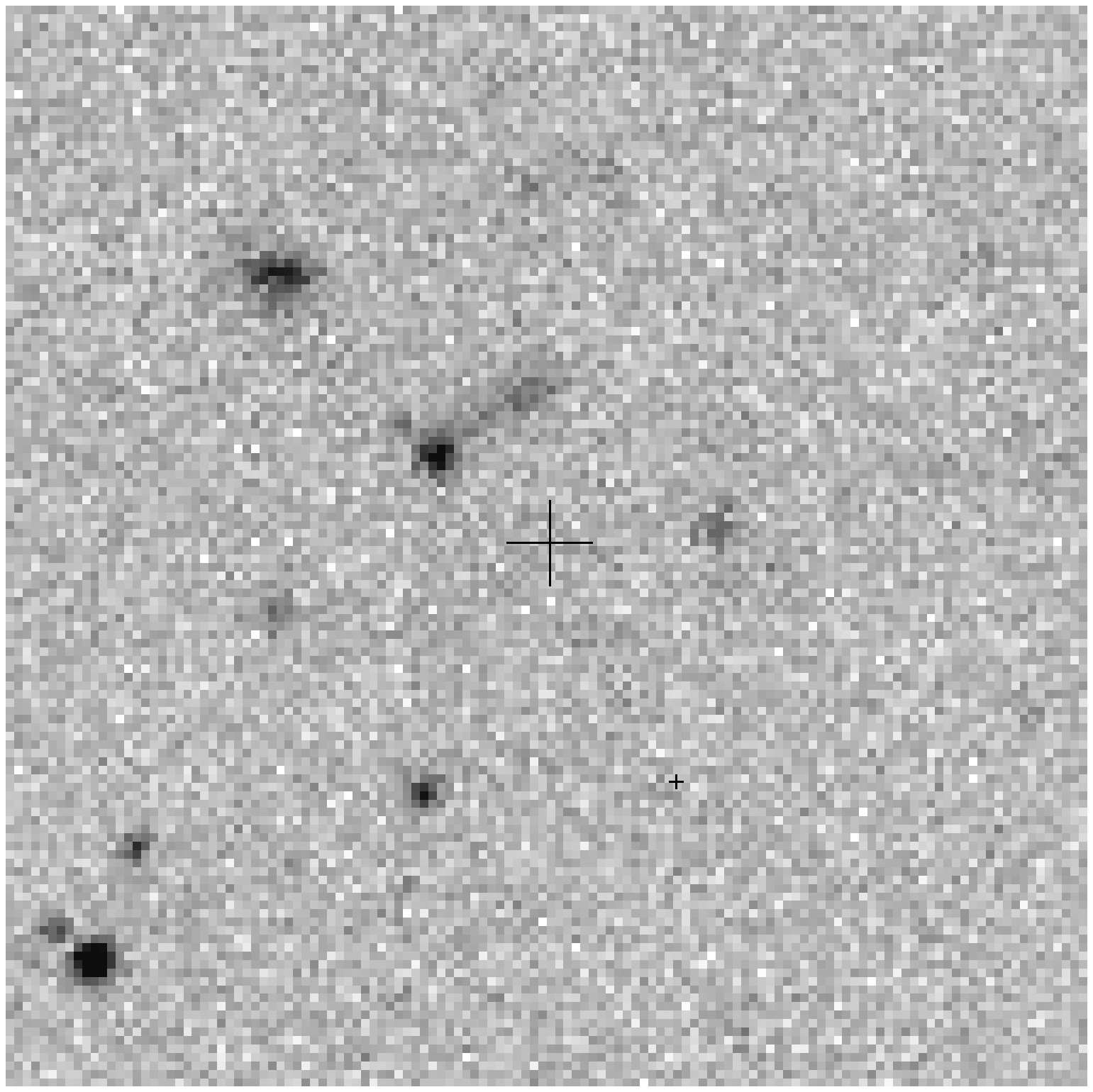}
\includegraphics[width=85mm]{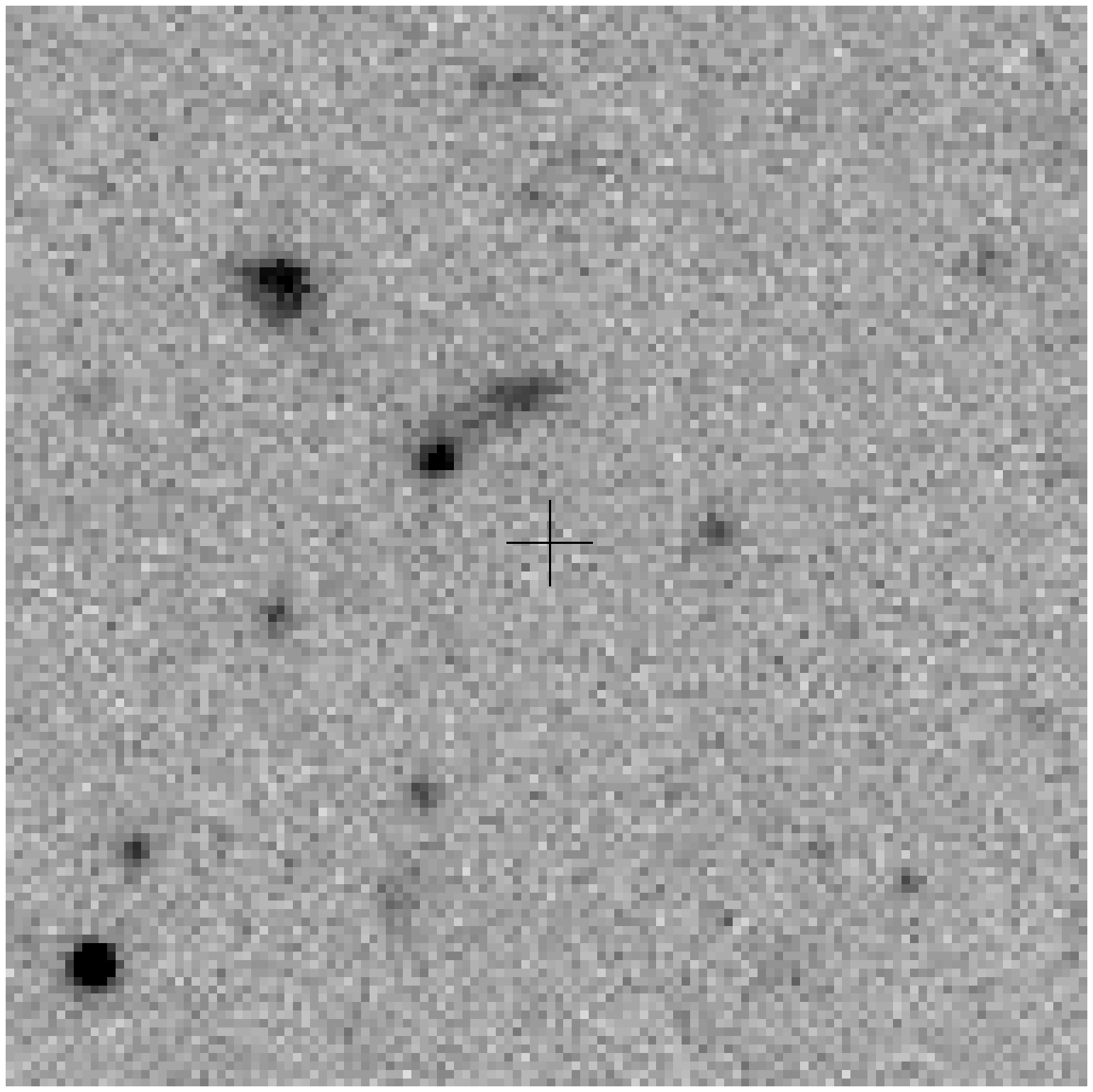}
\caption{From top left to bottom  left:  $U$, $B$ and $V$-band  images of the field
of the millisecond pulsar PSR J2124$-$3358
obtained with the  FORS1 instrument of the VLT-Antu. North to the top,
East to  the left. The images, $25" \times 25"$, are obtained by the
median of all the available exposures (see Table  1).  The cross marks
the nominal  radio position  of the pulsar  computed according  to the
more recent radio coordinates (see Gaensler, Jones and Stappers 2002). The sizes of the cross arms
is set to 1 arcsec, i.e., 5  times the overall  uncertainty on the pulsar  position
}  
\end{figure}

\noindent
Since  the error on  the pulsar position  is negligible,
the overall  uncertainty in the registration of  the radio coordinates
on the optical frames is  dominated by the accuracy of the astrometric
catalogue used as a reference and by the rms of the astrometric fit.
Taking  both  factors  into  account,  we finally  ended  up  with  an
uncertainty of  $\approx 0.2''$  in both RA  and Dec.   Any additional
uncertainty on the  pulsar position due to its  known proper motion of
$\mu=52.6$~mas~yr$^{-1}$   (Cordes,  Jones   and  Stappers   2002)  is
certainly within  our global  error budget.  We  note that  the GSC-II
astrometry was  tied to the  extragalactic radio source  frame (ICRF),
thus  our  position  should  not  be  significantly  affected  by  any
systematic  offset between  the radio  and optical  reference systems.
Our astrometry is shown on  the co-added $UBV$-band images (Figure 2).
No  object is  detected  at the  pulsar  position down  to $3  \sigma$
limiting magnitudes  of $U  \simeq 26$, $B\simeq  27.7$ and  $V \simeq
27.8$. Thus,  we assume  these values as  upper limits on  the optical
flux of  the pulsar.  By  scaling for the  assumed distance of  270 pc
(Cordes \& Lazio  2002), our upper limits imply  an optical luminosity
$L_{opt}  \le   2.5  ~  10^{26}$   ergs  s$^{-1}$,  which   makes  PSR
J2124$-$3358 intrinsically fainter than the old ($\approx 10^{6}-10^{7}$ yrs)
{\em  ordinary}  pulsars PSR  B1929+10  and  PSR  B0950+08 (Pavlov  et
al. 1996).

\section*{Conclusions}

We have reported on the first, deep, optical observations of the field
of the millisecond pulsars  PSR J2124$-$3358 performed with the ESO VLT. The pulsar is undetected
down to a flux limit of $V \sim 27.8$, the deepest obtained so far for
an object of  this class.  Unfortunately, this result  adds to the list
of  non-detections of  millisecond pulsars  deriving  from (shallower)
observations recently performed with the VLT, namely PSR J1024$-$0719,
PSR J1744$-$1134 (Sutaria et al.  2002) and PSR J0030+0451 (Koptsevich
et al. 2002).   Future observations of these objects in the near-UV  with the  HST will
probably offer higher chanches of detection.

%

\end{document}